\documentclass{aa}  

\usepackage{graphicx}
\usepackage{txfonts}
\usepackage{subcaption}
\usepackage{ulem}
\usepackage{placeins}
\usepackage{float}

\def\kms{km\,s$^{-1}$}

\def\vrad{${\rm V}_{r}$}

\def\He~I{${\rm He~I}$}

\def\Ca~II{Ca~II}

\begin{document}

   \title{Unraveling the binary nature of HQ Tau:}
   \subtitle{A brown dwarf companion revealed using multi-variate Gaussian process}

   \author{K.~Pouilly \inst{1}
          \and
          J. Bouvier \inst{2}
          \and
          E.~Alecian\inst{2}
          }

    \institute{Department of Astronomy, University of Geneva, Chemin Pegasi 51, CH-1290 Versoix, Switzerland\\
              \email{Kim.Pouilly@unige.ch}
              \and
              Univ. Grenoble Alpes, CNRS, IPAG, 38000 Grenoble, France}

   \date{Received September 25, 2023; accepted June 06, 2024}

% \abstract{}{}{}{}{} 
% 5 {} token are mandatory
 
  \abstract
   { Both the stellar activity and the accretion processes of young stellar objects can induce variations in their radial velocity (RV). 
   This variation is often modulated on the stellar rotation period and may hide a RV signal from a planetary or even a stellar companion.}
   {The aim of this study is to detect the companion of HQ Tau, the existence of which is suspected based on our previous study of this object. We also aim to derive the orbital elements of the system.}
   {We used multi-variate Gaussian process regression on the RV and the bisector inverse slope of a six-month high-resolution spectroscopic follow-up observation of the system to model the stellar activity. 
   This allowed us to extract the Keplerian RV modulation induced by the suspected companion.}
   {Our analysis yields the detection of a $\sim$50 M$_{jup}$ brown dwarf companion orbiting HQ Tau with a $\sim$126 day orbital period.
   Although this is consistent with the modulation seen on this dataset, it does not fit the measurements from our previous work three years earlier.
   In order to include these measurements in our analysis, we hypothesise the presence of a third component with orbital elements that are consistent with those of the secondary according to our previous analysis (M$_B\sim$48 M$_{jup}$, P$_{orb, B}\sim$126 days), and a $\sim$465 M$_{jup}$ tertiary with a $\sim$767 day orbital period.
   However, the hypothesis of a single companion with M$_B\sim$188 M$_{jup}$ and P$_{orb}\sim$247 days can fit both datasets and cannot be completely excluded at this stage of the analysis.}
   {At minima, HQ Tau is  a single-lined spectroscopic binary, and several factors indicate that the companion is a brown dwarf and that a third component is responsible for larger RV variation on a longer timescale.}


   \maketitle

%
%________________________________________________________________

\section{Introduction}

    The accretion processes of pre-main sequence (PMS) stars induce several forms of variability in stellar spectra.
    The classical T Tauri stars (CTTS) are low-mass PMS objects surrounded by an accretion disc and present a magnetically driven accretion process.
    The strong magnetic field of these objects exerts magnetic pressure on the disc, forcing the material to leave the disc plane and fall onto the star along the magnetic field lines, a process known as magnetospheric accretion \citep[see, \textit{e.g.},][as reviews]{Bouvier07b, Hartmann16}. 
    The material thus reaches the stellar surface in a very localised region near to the dipole magnetic pole.
    The material's kinetic energy is dissipated in a shock, heating the region and producing a bright spot, also called a hot or accretion spot.
    The signature of this spot on the stellar photospheric lines is the same as the cold spots, that is, the darker magnetic regions seen on the Sun, which are also present on CTTSs: a bump in the line profile, blueshifted and redshifted with the stellar rotation, and producing an apparent radial velocity (RV) modulation with the same period as the rotation period of the object \citep{Vogt83}.
    This feature is a well-known signature of stellar activity and may hide other modulations of the RV induced by a planetary or stellar companion.

    The aim of the present work is to study the RV of \object{HQ Tau} (right ascension (RA): 4h35m, declination (Dec):$+$22$^\circ$50'), an intermediate-mass T Tauri star (IMTTS) of which the magnetospheric accretion process was reported and studied in detail by \cite{Pouilly20} (hereafter Paper I).
    These latter authors reported a modulation of the RV on the star's rotation period and ascribed this to a large polar spot.
    However, the mean RV ($\langle$V$_r\rangle$=7.22$\pm$0.27 \kms) was far below the median velocity of the Taurus region \citep[about 15.5 \kms,][]{Galli19}, and was found to be inconsistent with the work of \cite{Nguyen12} and \cite{Pascucci15} (16.3$\pm$0.02 and 17.2$\pm$0.2 \kms, respectively).
    In addition to the erroneous renormalised unit weight error (RUWE) measured by \textit{Gaia} \citep[RUWE=13.39;][]{Gaia23}, we thus suspected the presence of a companion, and the additional RV measurements obtained in late 2019 and early 2020 confirmed a larger variation of the RV than that produced by the stellar activity only (see Table 3 of Paper I).

    HQ Tau was first suspected to be a tight binary by \cite{Simon87} based on lunar occultation, with a separation of 4.9$\pm$0.4 mas.
    This estimation was later revised by \cite{Chen90} to 9.0$\pm$0.2 mas, who noted that the binary nature of this system is not obvious.
    All subsequent studies \citep{Richichi94, Simon95, Simon96, Mason96} led to the same conclusion, that HQ Tau is a single object, until our work in Paper I.
    Indeed, the large RV variation, in addition to the inner cavity in HQ Tau's disc suspected by \cite{Long19} and \cite{Akeson19} from ALMA observations, are in favour of the presence of a companion, motivating further investigations of this aspect of this system.  

    The new dataset presented in this work covers a six-month period, which should allow us to probe the presence of a companion and derive the orbital elements of the system given the large RV variation that occurs over about 2 months (between our measurements in 2019 and 2020), meaning an orbital period of about 4 months if this variation corresponds to the peak-to-peak amplitude of the velocity modulation.
    The spectroscopic dataset is described in Sect.~\ref{sec:obs}, and the entire analysis is presented in Sect.~\ref{sec:results} and discussed in Sect.~\ref{sec:discussion}. 
    We present our conclusions from this work in Sect.~\ref{sec:conclusion}.

%__________________________________________________________________

\section{Observations}
\label{sec:obs}
    In this section, we describe the dataset used to perform our analysis. 
    We used four different instruments: the Spectropolarimètre Infrarouge \citep[SPIRou, ][]{Donati20b} and the Echelle SpectroPolarimetric Device for the Observation of Stars \citep[ESPaDOnS,][]{Donati03} at Canada-France-Hawaii Telescope (CFHT), Neo-Narval at Telescope \textit{Bernard Lyot} (TBL), and SOPHIE at Observatoire de Haute Provence (OHP). The 39 observations were performed in late 2020 and early 2021, covering almost 6 months in total, with a sampling cadence that varies between less than 1 day and 28 days. The complete log of observations is provided in Appendix~\ref{ap:logobs}.

    \subsection{SPIRou}

    Most of the observations taken at the CFHT were made using SPIRou, a near-infrared spectropolarimeter covering the YJHK band in a single shot at a resolution of about 75~000.
    The ten observations were taken between 2020 September 4 and 2021 January 5 using the polarimetric mode, meaning that they are composed of four subexposures in different polarimeter configurations, allowing us to derive the intensity (Stokes I), and the circularly polarised (Stokes V) and null spectra.
    The exposure time of each subexposure is about 150 s.
    The observations were reduced and telluric corrected using the \texttt{APERO} data reduction system \citep{Cook22}, and reached a signal-to-noise ratio (S/N) ranging between 73 and 90 for the subexposure combination at the order \#47.
    
    \subsection{ESPaDOnS}

    Four observations were performed using ESPaDOnS mounted at the CFHT between 2020 November 30 and 2020 December 8.
    ESPaDOnS covers the 300 to 1000 nm wavelength range and reaches a resolving power of 68~000. 
    Here again, observations were made using the polarimetric mode, with a $\sim$150 s exposure time for each subexposure.
    These observations were reduced using the \texttt{Libre-ESpRIT} package \citep{Donati97}, and reach a S/N of between 73 and 90 for the subexposure combination on Stokes I at 730 nm.
    Unfortunately, the observation on 2020 November 30, which has the highest S/N, seems to be contaminated by the Moon and is not used in this work. 
    
    \subsection{Neo-Narval}

    We obtained 11 observations at the TBL using the ESPaDOnS's twin Neo-Narval \citep{LopezAriste22} between 2020 September 5 and 2021 February 23. These data were reduced using the latest DRS version of the instrument.
    The polarimetric mode was used here again, with approximately 450 s of exposure time for each subexposure.
    The resolution of this instrument is about 65~000 and it covers the  380-1000 nm wavelength range.
    The S/N of the intensity spectrum ranges from 7 to 20 at 600 nm.
    
    \subsection{SOPHIE}

    The largest individual data set used in this work consists of 14 observations carried out with the SOPHIE spectrograph mounted at the OHP between 2020 September 3 and 2021 February 17. 
    The observations were taken in a single exposure of $\sim$1800 s.
    This instrument has a resolution of 40~000 and covers the 390 to 700 nm wavelength range.
    The raw data were reduced using the SOPHIE real-time pipeline \citep{Bouchy09}, and the resulting spectra  reach a S/N of between 40 and 64 at 600 nm.
    One of the observations (2020 November 27) was affected by Moon contamination and is therefore not used in this work.

%___________________________________________________________________    
    
\section{Results}
\label{sec:results}

    \subsection{Radial velocity}
    
\begin{figure*}
        \centering
        \sidecaption
        \includegraphics[width=.70\textwidth]{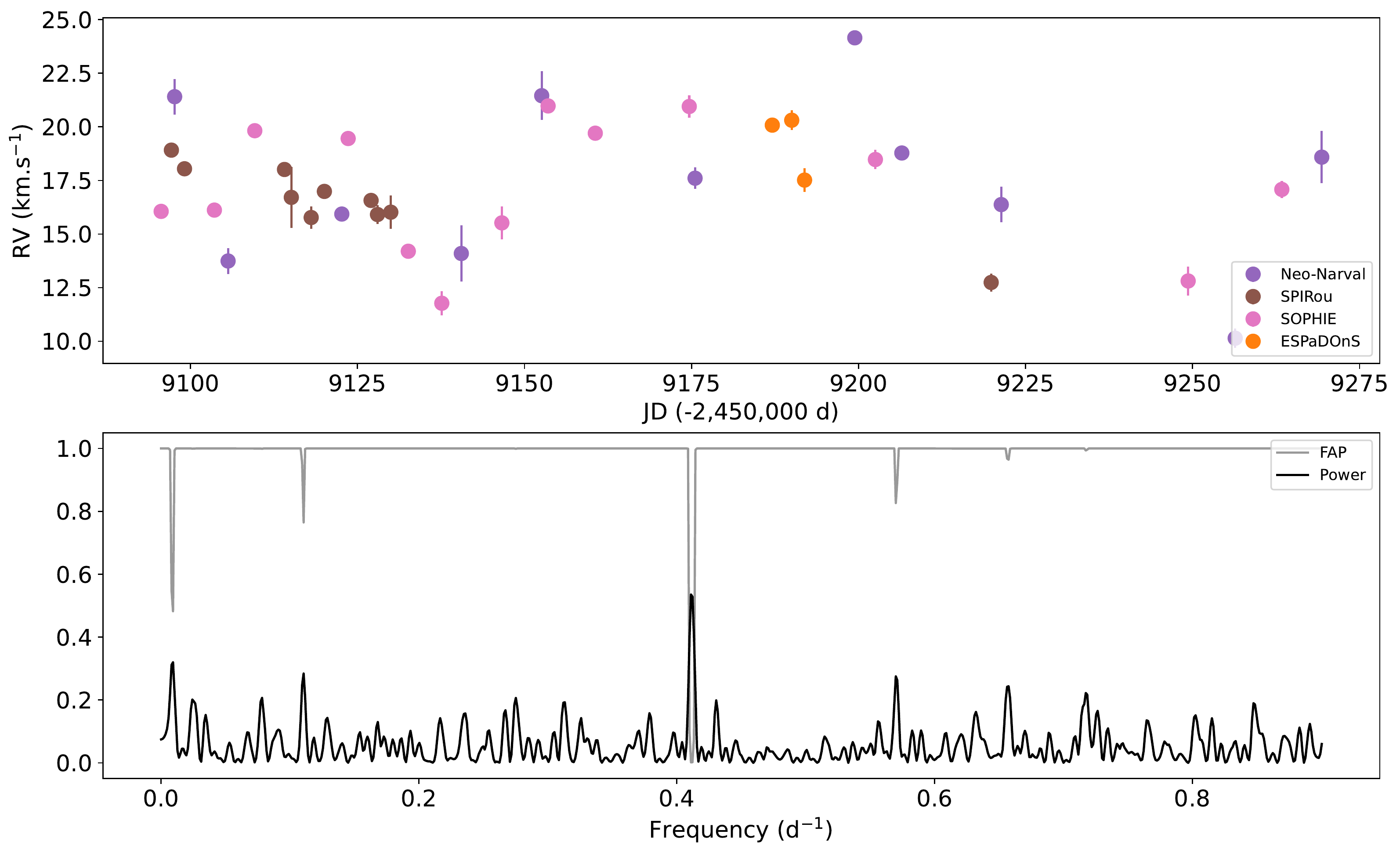}
        \caption{Radial velocity curve \textit{(top)} and corresponding Lomb-Scargle periodogram analysis \textit{(bottom)} of HQ Tau over the 20B semester. The colors of the radial velocity curve correpond to the different instruments used: NeoNarval at TBL (\textit{purple}), SOPHIE at OHP (\textit{pink}), ESPaDOnS (\textit{orange}), and SPIRou (\textit{brown}) at CFHT. On the bottom panel the Lomb-Scargle periodogram is shown in black and the FAP in grey.}
        \label{fig:RVHQTau}
    \end{figure*}

    The crucial first step of this analysis is to derive the RV.
    As described in Paper~I, the amplitude of the RV modulation seems to be of around 14 \kms, including a spot modulation of about 6 \kms.
    We are thus looking for an orbital modulation of the same order as the spot modulation.
    We used the cross-correlation method described in Paper~I to derive the RV using the same photospheric template (i.e. Melotte25-151 / \object{V1362 Ori}) for the observations in the visible frame, meaning the SOPHIE, Neo-Narval, and ESPaDOnS' spectra, and V819 Tau for the SPIRou observations.
    We computed the RV on ten wavelength windows of about 10 nm in width between 480 and 880 nm for each visible observation and on five windows, also of about 10 nm in width, between 1550 and 2150 nm for the SPIRou observations.
    Then, for each observation, we averaged the results of each window and computed a standard deviation to derive the RV and its uncertainty. All the computed values are provided in Table \ref{tab:vradCFHT} and the  \vrad\ curve is shown in Fig~\ref{fig:RVHQTau}.
    
    Although the values are relatively widely scattered, they follow a trend in that they oscillate around the mean of \vrad $\sim$ 17 \kms\ on a period of about 100 days.
    Furthermore, the amplitude of the variation (about 15 \kms) and the mean velocity are far larger than those measured in Paper~I ($\sim$~4~\kms\ and 7.22~\kms , respectively), while the photometric variations were within 0.14 mag in V band at the time this new dataset was obtained according to the American Association of Variable Star Observers (AAVSO) survey.
    This indicates that another phenomenon modulates the RV in addition to a surface spot.
    
    The Lomb-Scargle periodogram \citep{Scargle82} shown in Fig.~\ref{fig:RVHQTau} yields a detection at f$\sim$0.4 d$^{-1}$, with a false-alarm probability (FAP) of 10$^{-3}$ (computed following the approximation of \cite{Baluev08}), meaning P$\sim$2.5 d, which is consistent with a stellar rotation period, as expected. 
    Furthermore, we note a secondary peak at f$\sim$0.009 d$^{-1}$, meaning P$\sim$111 d, which is consistent with the apparent modulation seen on the RV curve; however, the FAP of 0.48 for this signal is too high for it to be considered as an independent detection of orbital motion.

    \subsection{Least-squares deconvolution profiles and bisector inverse slope}
    \label{subsec:LSD_BIS}

    The analysis of the bisector is commonly used to identify the origin of a RV variation.
    Indeed, when a companion shifts the line velocity, the stellar activity alters the line shape, producing an asymmetric line profile, which can be studied using the bisector inverse slope \citep[BIS;][]{Queloz01}.
    
\begin{figure*}
        \centering
        \includegraphics[width=.99\textwidth]{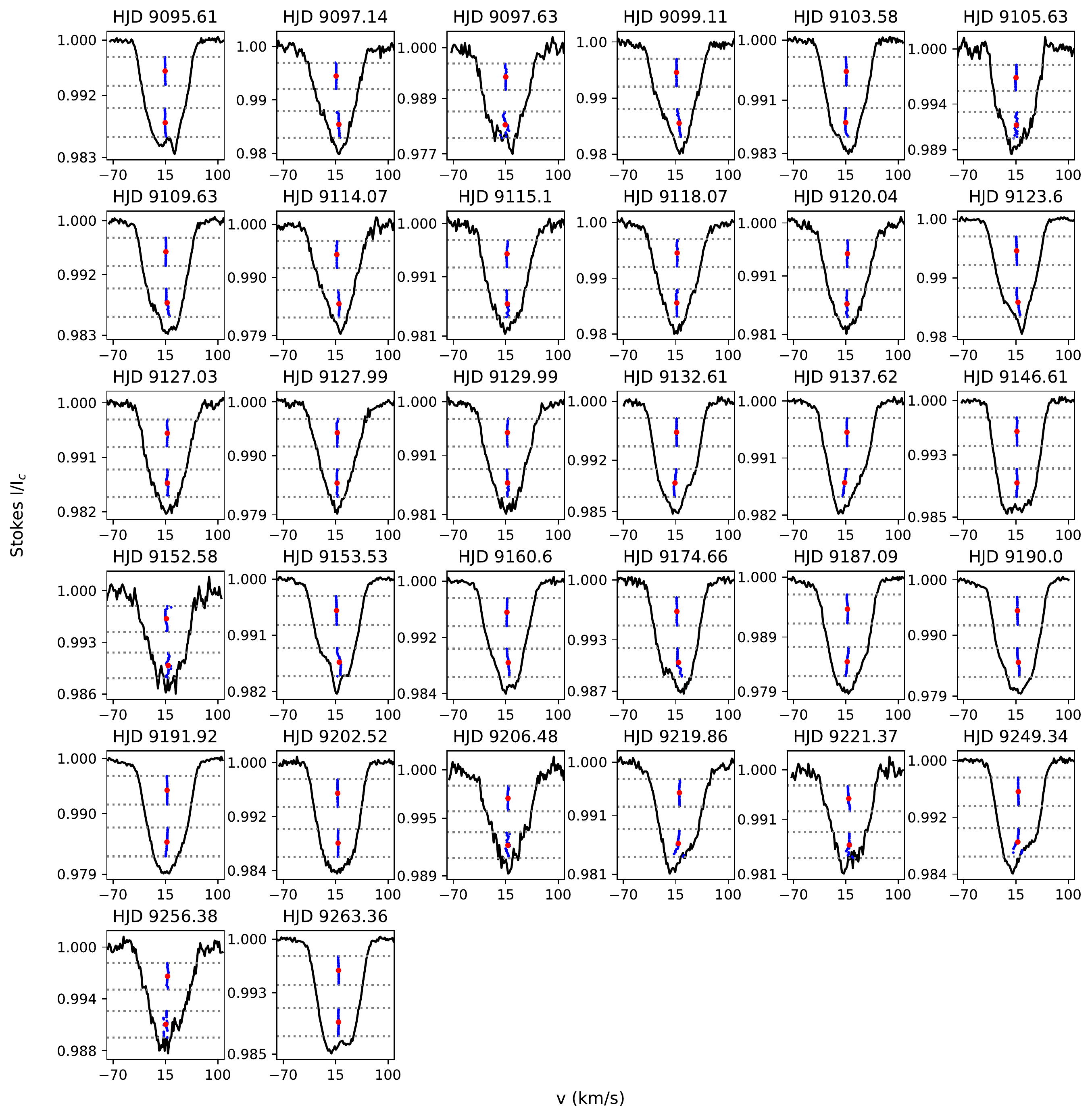}
        \caption{Measurement of the BIS of the Stokes I profiles. The HJD is indicated at the top of each plot. The black lines are the Stokes I profiles and the grey dotted line delimits the top and bottom regions selected for the BIS computation. The blue points are the bisector in these regions and the red dots are the mean bisector of each region used to compute the BIS. }
        \label{fig:bis}
    \end{figure*}

    Here, we study the BIS of the least-squares deconvolution \citep[LSD;][]{Donati97} profiles of the unpolarised signal (Stokes I).
    The S/N increase provided by this method, as well as the switch to the velocity space, allow us to take into account the observations with low S/N, as well as their different wavelength regions.
    In order to normalise the LSD weights, we used an intrinsic depth of 0.2, a mean Landé factor of 1.2, and a mean wavelength of 520 (1200) nm for the optical (infrared) spectra.
    For the computation, we used a mask derived from a line list provided by the VALD \citep{Ryabchikova15} database covering the wavelength range of each instrument and from which we excluded the regions contaminated by emission lines or telluric absorption.
    The resulting Stokes I profiles used in this study are shown in Fig.~\ref{fig:bis}.
    
    To compute the BIS, we selected two regions at the top and the bottom of the line, in which we computed the mean velocity of the bisector.
    The BIS is then defined by the difference between the top velocity and the bottom one.
    The two regions were selected as follows: we ignored the first 15\% of the profile containing the continuum and the wings of the line, as well as the last 15\% containing the bottom of the profile, where the bisector cannot be computed because of the noise or because of an excessively strong activity signature splitting the profile into two parts in this region.
    The top region is thus defined as 25\% of the profile from the upper limit, and the bottom region is defined as 25\% of the line from the lower limit.
    The error on the BIS is defined as the quadratic mean of the standard deviation of the bisector of the top and bottom regions.
    We removed five observations\footnote{HJD (-2\,450\,000 days) 9122.65, 9140.54, 9175.53, 9199.44, and 9269.34.} from this analysis because the S/N did not allow us to properly derive the bisector of the profile. 
    The computation of the BIS is illustrated in Fig.~\ref{fig:bis}, and the values are provided in Table~\ref{tab:vradCFHT}.
    
\begin{figure}
        \centering
        \includegraphics[width=.45\textwidth]{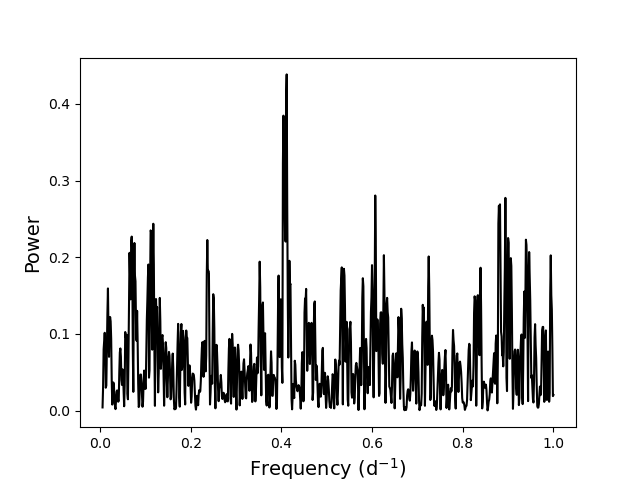}
        \caption{Lomb-Scargle periodogram of the BIS. The peak at f=0.41 d$^{-1}$ corresponds to a period of 2.42 d and has a FAP of 0.09.}
        \label{fig:BISperiodo}
    \end{figure}

\begin{figure}
        \centering
        \includegraphics[width=.45\textwidth]{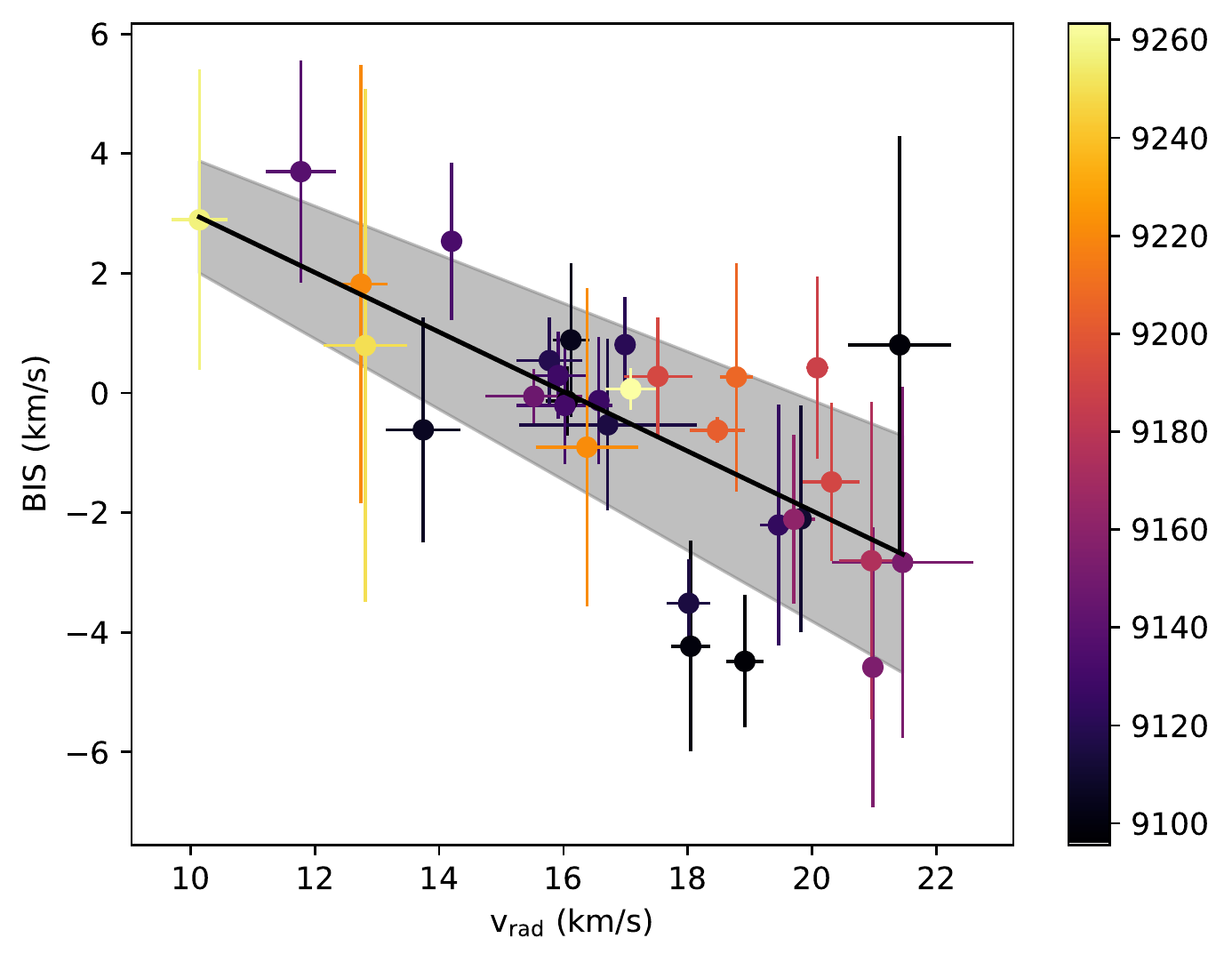}
        \caption{BIS versus RV. The colour scale corresponds to the HJD, the black line is the linear fit, and the grey-shaded area is the 1$\sigma$ uncertainty on the slope of the fit.}
        \label{fig:RVxBIS}
    \end{figure}

    The BIS varies with a period of 2.42 days (f=0.41 d$^{-1}$, FAP=0.09; see Fig.~\ref{fig:BISperiodo}), which is consistent with the rotation period of the star and thus confirms that it is tracing the stellar activity. 
    We do not observed any signal around 111 d. 
    Figure~\ref{fig:RVxBIS}  shows the BIS versus the RV.
    In the case of a RV that is fully modulated by the stellar activity, a linear relation with a slope of -1 is expected, and a BIS of 0 \kms\ is expected at the star's real velocity as well.
    For HQ Tau, a linear downward trend is observed, as well as a clustering of measurements around BIS~=~0~\kms\ and \vrad $\sim$ ~17~\kms\ (the mean velocity of our data set).
    However, the fitted slope is $-$0.50 $\pm$ 0.09, and the measurements show a larger dispersion at the lower and higher velocities. 
    This indicates that, in addition to the stellar activity, another phenomenon is affecting the RV modulation.

    \subsection{Multi-dimensional Gaussian process regression}
    \label{subsec:mutliGPR}
    
    Although visual inspection of the RV curve and the Lomb-Scargle periodogram analysis seems to indicate the presence of a second modulation with a lower frequency and lower amplitude than the stellar activity, the signal remains insufficiently strong to constrain any orbital parameters.
    We therefore used Gaussian process regression (GPR) to model the stellar activity signal with \texttt{pyaneti}, a tool developed and fully described by \cite{Barragan19}; see also \citet[][]{Barragan22} for the latest version.
    Here we simply describe its basic principles.
    This method, which is often used in searches for exoplanets around active hosts, considers that the modulation can be described as a finite set of random variables related by a multi-variate normal distribution, defining a Gaussian process (GP).
    According to \cite{Tracey18}, this distribution is given by:

    \begin{equation}
        P(\mathbf{t}) = \frac{1}{\sqrt{(2\pi)^{N}|\mathbf{K}|}} \mathrm{exp}\left[ -\frac{1}{2} (\mathbf{t}-\boldsymbol{\mu})^{T} K^{-1} (\mathbf{t}-\boldsymbol{\mu}) \right],
    \label{eq:GPdistrib}
    \end{equation}
    
    \noindent where \textbf{t} = t$_{i,(i=1,...,N)}$ are the locations of the N variables, $\boldsymbol{\mu}$ are the mean values, and \textbf{K} is the covariance matrix (or kernel), giving the relation between the variables.

    The aim of GPR is, assuming that the time series is a sample of GP, to optimise the parameters $\boldsymbol{\phi}$ of the mean function $\mu(t;\boldsymbol{\phi})$, and the hyper-parameters $\boldsymbol{\Phi}$ of the kernel function $\gamma(t_i,t_j:\boldsymbol{\Phi})$.
    In our case, $\mu$ is physically motivated by a Keplerian, and $\gamma$ is a composition of the `white-noise' (WN) and `quasi-periodic' (QP) kernels, to describe the rotational modulation and the error of our data.
    These kernels are given by

    \begin{equation}
        \gamma_{WN}(t_i,t_j)=\sigma^2_i \delta_{ij}
    ,\end{equation}
    where $\sigma_i$ is the error of the value $i$, and $\delta_{ij}$ is the Kroenecker delta, and
    \begin{equation}
        \gamma_{QP}(t_i,t_j) = A^2 \mathrm{exp} \left\{ -\frac{\sin^2 \left[ \pi (t_i - t_j)/ P_{GP} \right]}{2 \lambda^2_p} - \frac{(t_i - t_j)^2}{2 \lambda^2_e} \right\},
    \end{equation}
    where $A$ is the amplitude scaling the deviation from the mean function, $P_{GP}$ is the period of the GP (in our case, the rotation period), $\lambda_p$ is the inverse of the harmonic complexity (i.e. the complexity of the variation within a period), and $\lambda_e$ is the long-term evolution timescale.

    In this work, we used the multi-dimensional GPR implemented in \texttt{pyaneti}, meaning that we used two time series simultaneously: the RVs and an activity indicator, in our case the BIS of the LSD profiles.
    These two sets of variables are described by a composition of the function drawn by the GP, $G(t)$, and its derivative $\dot{G}(t)$ as follows:

    \begin{equation}
    \begin{cases}
        \Delta RV(t) = A_1 G(t) + B_1\dot{G}(t) \\
        BIS(t) = A_2 G(t) + B_2\dot{G}(t),
    \end{cases}
    \end{equation}
    \noindent where $\Delta RV$ is the RV modulation due to both the stellar activity and the companion. 
    The RV is described by RV(t) = RV$_{syst}$ + $\Delta RV$, where RV$_{syst}$ is the systemic RV.
    
    Physically, the function $G(t)$ represents the projected area of the active regions on the visible stellar surface that induces the RV (and other activity indicator) variations, as a function of time.
    The RVs and the BIS are affected by these regions, and by their time evolution \citep{Dumusque14}; the best description of these data is thus given as a function of $G(t)$ and its time derivative $\dot{G}(t)$.

\begin{figure}
        \centering
        \includegraphics[width=0.49\textwidth]{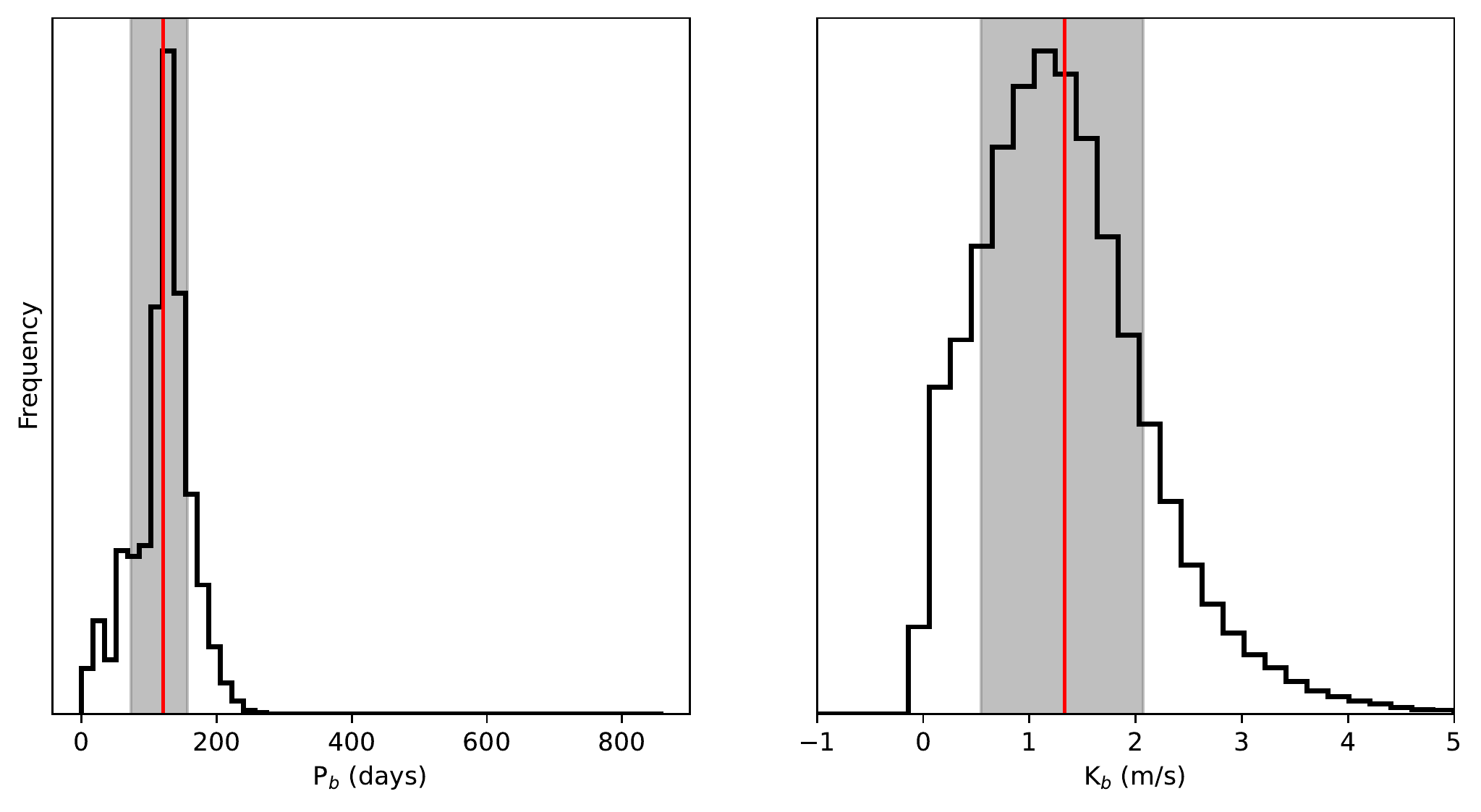}
        \caption{Posterior distribution of the orbital period \textit{(left)} and of the semi-amplitude \textit{(right)}. The vertical red line shows the mean value and the grey-shaded area illustrates the 68.8\% confidence level.}
        \label{fig:GPposterior}
    \end{figure}

\begin{figure*}
        \centering
        \sidecaption
        \includegraphics[width=0.70\textwidth]{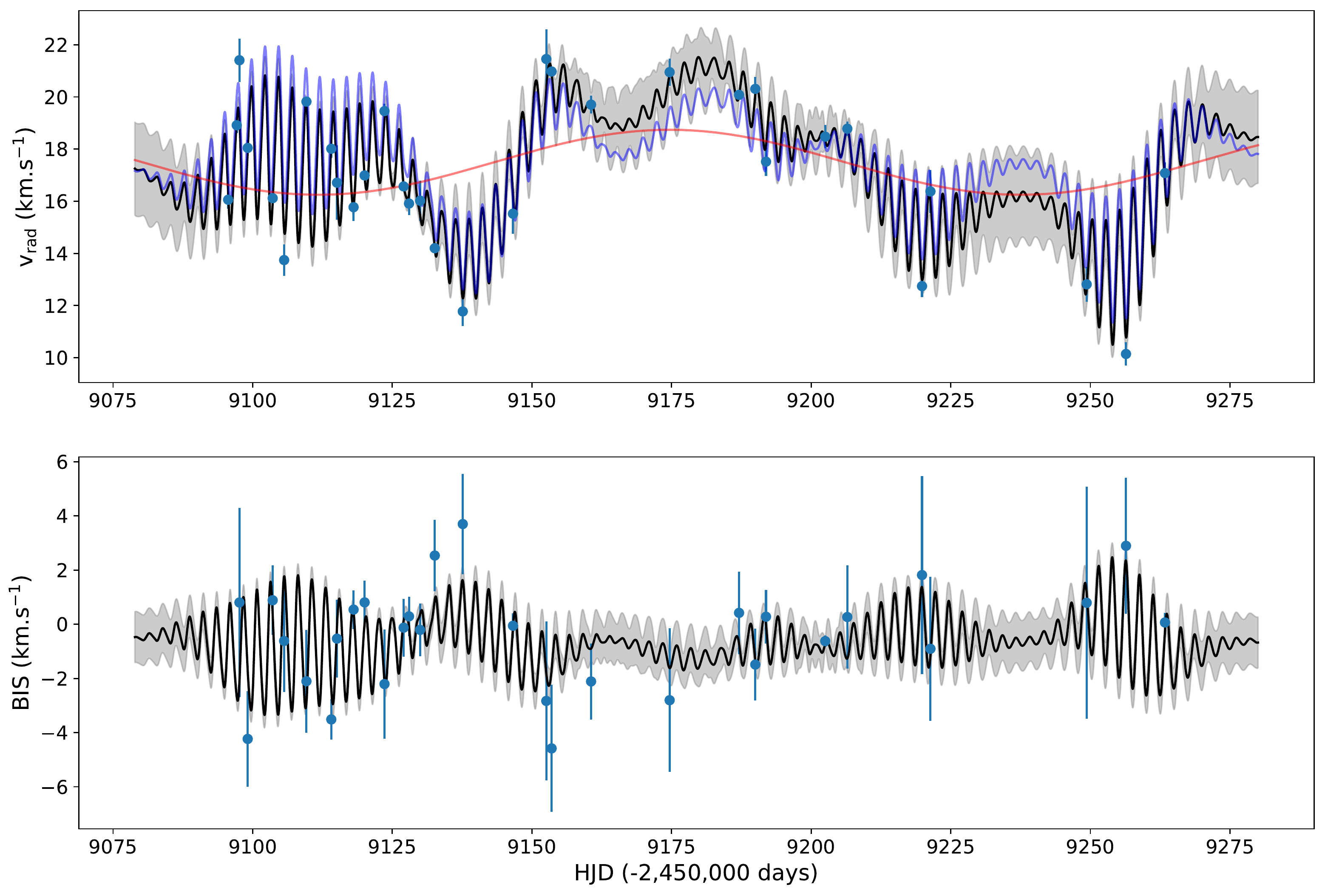}
        \caption{Fit of the RV \textit{(top)} and BIS \textit{(bottom)} curves with \texttt{pyaneti}. The blue points are the measurements with their uncertainties. On the top panel the black curve is the GP fit of the stellar activity \textit{(purple)} plus the Keplerian modulation by the companion \textit{(red)}. The grey-shaded area is the 1$\sigma$ uncertainty of the fit.  On the bottom panel the black curve is showing the GP fit of the stellar activity only with its 1$\sigma$ uncertainty in grey.}
        \label{fig:GPts}
    \end{figure*}

    The optimisation of the complete set of parameters is done through a Markov Chain Monte Carlo (MCMC) procedure using 500 chains to sample the parameter space.
    To perform the simultaneous fit of RV and BIS, we need both values for each observation; we therefore had to exclude the RV measurements for which we could not measure the BIS from the fit (see Sect.\ref{subsec:LSD_BIS}).
    This analysis yielded the detection of an orbital modulation in the RVs with a semi-amplitude of K~=~1.25$^{+0.83}_{-0.71}$~\kms\ and a period of P$_{orb}$~=~125.8$^{+31.5}_{-55.2}$ d, in addition to a RV modulation by stellar activity consistent with the period of Paper I, and a systemic velocity of RV$_{syst}$~=~17.5$^{+0.9}_{-1.0}$~\kms, which is consistent with the median velocity of the Taurus cloud \citep[v~=~16.4$\pm$1.1~\kms,][]{Galli19}. 
    Using the binary mass function and the Kepler's third law, this means that a companion, a brown dwarf with a minimal mass of 50.1$^{+29.6}_{-32.0}$ M$_{jup}$, is orbiting HQ Tau with a semi-major axis of about 0.61 AU.
    The complete set of parameters inferred by this analysis and the prior used are provided in Table~\ref{tab:GPresults}.
    In addition, the posterior distribution of the orbital period and semi-amplitude is shown in Fig.~\ref{fig:GPposterior}, and the resulting curves for RV and BIS are shown in Fig.~\ref{fig:GPts}.
    We did not succeed in constraining the orbit's eccentricity and the argument at periastron (we obtained flat posterior distributions), and therefore we assume a circular orbit and $\omega$=0.
    The hyperparameters of the GP are consistent with expectations, meaning that the GP's period recovers the stellar rotation period, the $A_1$ and $A_2$ coefficients are of opposite sign, indicating a phase opposition of the RV and BIS modulations, and $\lambda_e$ (representing the spot's lifetime) is larger than the rotation period, indicating that this activity region can be used to recover the rotation period.

    \begin{table}
        
        \caption{Parameters of HQ Tau B.}
        \centering
        \begin{tabular}{l c c}
        \hline\hline
            Parameter & Prior & Value \\
            \hline
            
            Fitted parameters & & \\
            T0 (HJD-2\,450\,000)& $\mathcal{N}$[9095,40]  & 9080.27$^{+21.25}_{-20.41}$ \\
            P$_{orb}$ (d) & $\mathcal{N}$[100,50] & 125.8$^{+31.5}_{-55.2}$ \\
            e & $\mathcal{F}$[0] &  \\
            $\omega$ & $\mathcal{F}$[0] &  \\
            K (\kms) & $\mathcal{U}$[-5.0, 5.0] & 1.25$^{+0.83}_{-0.71}$ \\
            
            A$_1$ (\kms) & $\mathcal{U}$[-10.0,0.0] & -1.50$^{+4.63}_{-2.16}$ \\
            B$_1$ (\kms) & $\mathcal{U}$[-10.0,10.0] & -1.81$^{+5.89}_{-2.76}$ \\
            A$_2$ (\kms) & $\mathcal{U}$[-10.0,10.0] & 0.33$^{+1.41}_{-1.75}$ \\
            B$_2$ (\kms) & $\mathcal{U}$[-5.0,5.0] & 1.63$^{+2.29}_{-5.16}$ \\
            $\lambda_e$ (d) & $\mathcal{U}$[1.,30] & 7.51$^{+2.19}_{-1.89}$ \\
            $\lambda_p$ & $\mathcal{U}$[0.1,5] & 2.48$^{+1.14}_{-0.85}$ \\
            P$_{GP}$ (d) & $\mathcal{N}$[2.42, 0.05] & 2.43$^{+0.04}_{-0.04}$ \\
            \hline
            %&&\\
            Derived parameters & & \\
            RV$_{syst}$ (\kms) &  & 17.5$^{+0.9}_{-1.0}$ \\
            %BIS$_{syst}$ (\kms) &  & -0.6$^{+0.6}_{-0.5}$ \\
            M$_B$ (M$_{jup}$)&  & 50.1$^{+29.6}_{-32.0}$ \\
            T$_{peri}$ (HJD-2,450,000) &  & 9049.30$^{+28.57}_{-26.65}$ \\
            a (AU) & & 0.61$^{+0.13}_{-0.24}$ \\
            %&&\\
            \hline
        \end{tabular}
        \tablefoot{$\mathcal{U}[a,b]$ represent a uniform prior between $a$ and $b$, $\mathcal{N}[a,b]$ a normal prior with a median $a$ and standard deviation $b$, and $\mathcal{F}[a]$ a value fixed at $a$. The uncertainty represents the 68.3\% confidence level of the posterior distribution. Please note that the mass M$_B$ corresponds to the minimal mass M$_B$$\sin$(i), where i is the orbit inclination.}
        \label{tab:GPresults}
    \end{table}

In order to decipher whether or not the stellar activity alone can explain the observed variability, we performed the same analysis, but this removing the Keplerian modulation.
The results are presented in Appendix~\ref{ap:GPonly}.
We then compared the Akaike information criterion \citep[AIC;][]{Akaike74} in its corrected prescription \citep[AICc;][]{Sugiura78} for a small number of data points; the AICc estimates the quality of a given model compared to another one through an estimate of the amount of information lost by the model.
The AICc obtained for the model including the Keplerian modulation is 304.77, while that obtained without the Keplerian modulation is 322.07, indicating that the model that includes the Keplerian modulation is preferred.
Finally, we performed a leave-one-out cross validation (LOOCV) test to assess the robustness of this solution.
This test consists of performing the analysis including all data points except one and computing the mean squared error (MSE) of the model at this point. 
This is done as many times as the number of data points, excluding a different measure each time. 
The final MSE is then derived as the average of all the individual MSEs computed.
The MSEs obtained for our solution are 6.96 and 3.41 \kms\ for the RV and the BIS, respectively.
The corresponding values for the model assuming only the stellar activity are 7.16 and 3.62 \kms, confirming that the Keplerian needs to be included to explain the observed variability.
Together with the MSE, we obtain the Keplerian parameters, in particular T0 and K, which yield HJD 2\,459\,081.42$\pm$4.83 and 1.35$\pm$0.20~\kms, respectively.
These values are perfectly consistent with the parameters of Table~\ref{tab:GPresults}, demonstrating the stability and consistency of this solution.

%______________________________________________________________

\section{Discussion}
\label{sec:discussion}

    HQ Tau was selected for this semester-long RV follow-up because of the surprising intermediate results of Paper I on a two-week high-resolution spectropolarimetric 2017 time series.
    In order to study the activity of this object, we derived the RV curve, from which we draw the  conclusion that a surface spot is modulating the RV on the stellar period.
    However, the mean RV is not consistent with previous measurements in the literature, and the curve shows a slight downward trend over the time series. 
    Two additional snapshot observations taken in late 2019 and early 2020 reveal a larger velocity modulation around the already published values, suggesting an additional source of RV modulation superimposed on the activity signal, presumably a companion.

    To disentangle the RV signal of the activity from the modulation induced by the suspected companion, we performed a multi-variate GPR using the Python tool \texttt{pyaneti} \citep{Barragan19}, reproducing the activity signal of the BIS and RV modulation with a GP, and adding a Keplerian signal to the latter.
    The MCMC procedure to recover the Keplerian's parameters and GP hyperparameters yields a Keplerian period of P$_{\rm orb}$ = 125.8$^{+31.5}_{-55.2}$~d, consistent with the Lomb-Scargle periodogram of the RV curve, and a semi-amplitude of K~=~1.25$^{+0.83}_{-0.71}$~\kms.
    Together with the activity-induced modulation reported in Paper I ($\sim$6~\kms\ peak-to-peak), this yields a peak-to-peak amplitude of about 10~\kms, consistent with the dataset's peak-to-peak variability ($\sim$13~\kms).
    Furthermore, the derived time at periastron and the fitted orbital parameters yield a minimum at about HJD 2~458~103, which is consistent with the downward trend of the RV modulation observed in Paper I (between HJD 2~458~055 and HJD 2~458~067).
    
    Concerning the stellar activity, the period of the GP, which is expected to be the stellar rotation period in our case, is perfectly consistent with our previous results and the Lomb-Scargle periodogram of the present dataset as well, and the $\lambda_e$>P$_{GP}$ confirms that this period is relevant and that the quasi-periodic kernel can be used for this analysis.
    % The harmonic complexity is relatively high (\textit{i.e.,} $\lambda_p \lesssim$ 1), which relates the moderate-to-high activity of the star, as expected for spot-dominated and fast-rotating young stars \citep[see the work on K2-100b by][]{Barragan19b}, which is the case of HQ Tau (see Paper I).
    The harmonic complexity is relatively low ({i.e.} $\lambda_p \gtrsim$ 1), which indicates a moderate to low activity of the star, as expected from the more evolved status of HQ Tau, and the close-to-sinusoidal modulation of its RV (see Paper I).

    These orbital parameters, together with the stellar parameters, yield a semi-major axis of 0.61~AU, which is fully consistent with the first estimation of \cite{Simon87} (4.9~mas separation; {i.e.} 0.78~AU at the distance of HQ Tau), and is inside the inner disc. 
    The presence of such a companion within the inner disc could possibly disrupt it \citep[][en reference therein]{Long18}.
    As discussed in Paper I, this is consistent with the ALMA observation modelled by \cite{Long19}, where the authors highlight dust depletion towards the inner disc region, indicating the presence of a dust cavity, which is nevertheless not well resolved in their data.
    Assuming that this dust cavity is produced by the companion, this means that the orbit inclination is the disc inclination (i.e. 53.8$^\circ$), constraining the companion's mass to about 62~M$_{jup}$.% without significant change of the orbit semi-major axis.

    Although the results are consistent with the present dataset, the
derived semi-amplitude does not allow us to reach the mean velocity obtained in Paper I.
    There are several possible explanations for this. Firstly, this six-month follow-up analysis may not be representative, given the three-year gap between the two datasets.
    The  derived $\lambda_e$ of 7.51 days, although larger than the rotation period as needed for the quasi-period kernel used, remains short regarding the time range of the studied phenomenon.
    This short spot lifetime indicates that the activity signal might change quickly, especially between 2017 and 2020, where the modulation by the stellar activity might have changed.
    Long-term RV variations may occur in active stars due to magnetic cycles that change the cool spot configuration at the stellar surface \citep{Grankin08}, or due to a slowly evolving surface magnetic field that modifies the distribution of surface hot spots \citep[e.g. see the multi-epoch studies of GQ Lup, DN Tau, and V2129 Oph by ][respectively]{Donati11, Donati12, Donati13}.
    However, in all cases, the amplitude of activity-related RV variations is expected to amount to a fraction of the star's v$\sin$i and, more importantly, the stellar RV averaged over a rotational cycle should vary very little as the spot modulation induces nearly symmetrical RV variations relative to the star's rest velocity. 
    This is not what is observed for HQ Tau, where the rotationally averaged RV in 2017 strongly departs from that observed in 2020/2021.
    A second explanation could be the presence of a second companion; with a much longer period, the RV modulation cannot be observed in the measurements analysed here, but could yield significantly lower velocities three years earlier.

    Aiming to begin the exploration of these hypotheses, we investigated the multi-variate GPR using \texttt{pyaneti} on the combined 2017 and 2020/2021 data sets.
    We performed the same analysis as in Sect.~\ref{subsec:mutliGPR}, in two configurations for the two hypotheses, meaning one or two companions modulating the RV curve in addition to the stellar activity, and the results are shown in Appendix~\ref{ap:GPR1720}.
    The results from the one-companion hypothesis appear inconsistent with the velocities reached in the 2017 measurements.
    The velocity modulation induced by the companion is still about 5~\kms\ above that measured, which is compensated for by a global decrease in the modulation by the stellar activity.
    Furthermore, the orbital period obtained ($\sim$250d) is inconsistent with the periodogram analysis.
    However, the systemic velocity derived (RV$_{syst}$~=~15.5$\pm$1.5~\kms) is consistent with the median velocity of the Taurus cloud \citep{Galli19}.
    The two-companion hypothesis yields consistent results with both the analysis on the 2020 data set only and the measurements of Paper I.
    HQ Tau B would be a 47.7~M$_{jup}$ brown dwarf, orbiting with a period of 126 d, which is consistent at 1$\sigma$ with the results of Sect.~\ref{subsec:mutliGPR}.
    HQ Tau C would exhibit a much larger mass (465~M$_{jup}$) and be on a 767 d orbital period (meaning a 2.16 AU semi-major axis), allowing it to reach the velocity determined in 2017, with the corresponding downward trend making its influence almost invisible during the six-month 2020 measurements.
    The systemic velocity resulting from this fit, of RV$_{syst}$~=~12.1$^{+1.4}_{1.8}$~\kms,\, is consistent at only 2$\sigma$ with the median velocity of the Taurus cloud derived by \cite{Galli19}.

    We performed the same test as in Sect.~\ref{subsec:mutliGPR} to assess the robustness of the different solutions, that is, we compared the AICc of the regression and the MSE from a LOOCV test.
    For completeness, we also performed the multi-variate GPR assuming the stellar activity only (see Appendix~\ref{ap:GPR1720}) and included it in the comparison.
    The results are summarised in Table~\ref{tab:1720test}, and are all in favour of the two-companion hypothesis.

    \begin{table}
        \caption{Results of the statistical tests performed for the joint fit of the 2017 and 2020 data sets.}
            \centering
            \begin{tabular}{lllll}
                \hline
                \hline
                Comp. & AICc & MSE$_{\rm RV}$ & MSE$_{\rm BIS}$ & T0/K \\
                \# && (\kms) & (\kms) & (HJD)/(\kms)\\
                \hline
                0 & 380.01 & 6.29 & 2.99 & \\
                1 & 379.47 & 6.12 & 2.83 & T0=9003.0$\pm$3.8\\
                &&&& K=3.9$\pm$0.3\\
                2  & 373.54 & 5.54 & 2.80 & T0$_{\rm B}$=9085.5$\pm$2.0\\
                &&&& K$_{\rm B}$=1.3$\pm$0.1\\
                &&&& T0$_{\rm C}$=9385.5$\pm$11.3\\
                &&&& K$_{\rm C}$=5.9$\pm$0.1\\
                \hline
            \end{tabular}
            \tablefoot{The colums are listing the number of companions assumed, AICc of the multi-variate GPR, the MSE from the LOOCV test on the RV and BIS, and the Keperian parameters T0 and K from the LOOCV test.}
            \label{tab:1720test}
        \end{table}
    
    %Corresponding semi-major axis 2.91 AU

    In addition to a good fit of all our RV measurements, the orbital period of the tertiary is close to the occurrence of the long-lasting UX Orionis events reported by \cite{Rodriguez17}, which take place  every 700 to 1000 days. One such event occurred immediately before the observations studied in Paper~I,
    and we discussed them therein, putting forward possible interpretations, such as a change of the magnetospheric radius relative to the dust sublimation radius, producing dipper-like occultations, or as a sudden change of the disc's vertical scale height, occulting the central star.
    This latter change in scale height could result from a disc wrapped by this component.
    As modelled by \cite{Cuello19, Cuello20}, stellar inclined retrograde flybys efficiently produce such a warp, in addition to a tilt of the disc, which remains after the warp dissipation, and could explain the discrepancy between the inclination of the stellar rotation axis ($\sim$75$^\circ$, Paper I) and the disc inclination \citep[53.8$^\circ$;][]{Long19}.
    
    To fully confirm the detection of such a tertiary companion, a set of observations similar to our 2020 dataset outside the inflexion point of the tertiary orbital motion is needed.
    Infrared high-resolution spectroscopy observations with a significantly
higher S/N than the present dataset might also allow the direct detection of the tertiary's spectra.
    Finally, given the mass(es) of the companion(s) and/or the semi-major axis of its (their) orbit(s), we assume that their influence on the  accretion process of the primary studied in Paper I can be neglected.
    Even if a flyby by the hypothetical tertiary were able to induce mass-accretion-rate enhancements \citep{Cuello19}, they only occur at low orbital inclination, contrary to the disc-warping phenomenon, and the variation is so drastic that the values derived in Paper I would have been clearly affected if HQ Tau had been observed during this phase.

%______________________________________________________________

\section{Conclusions}
\label{sec:conclusion}
    
    In this paper, we present an investigation of the companion hypothesis to explain the low RVs measured for  HQ Tau in our previous work \citep[][Paper I]{Pouilly20}.
    We monitored the RV of  HQ Tau over 6 months using high-resolution spectroscopy from four instruments in optical and IR frames (ESPaDOnS and SPIRou at CFHT, Neo-NARVAL at TBL, and SOPHIE at OHP).
    Finally, we modelled the stellar activity signal from the RV and BIS of the LSD profiles using multi-dimensional GPR to extract the Keplerian RV signal thanks to the \texttt{pyaneti} tool \citep{Barragan19, Barragan22}.

    This procedure yields the detection of a brown dwarf companion HQ Tau B (M$_{B}$=50$^{+30}_{-32}$ M$_{jup}$) orbiting the primary on a period of 126$^{+32}_{-55}$ days without affecting its spectrum.
    Although this companion is consistent with the lunar occultation detection of \cite{Simon87} and with the inner cavity in the disc suspected by \cite{Akeson19}, it does not allow us to recover the velocities measured in Paper I.

    To address this issue, we investigated the possible presence of a third component and performed the same analysis including both datasets.
    This yielded a secondary component showing parameters consistent we the previous analysis and a tertiary of a mass of 465$^{+186}_{-106}$ M$_{jup}$ and an orbital period of 767$^{+91}_{-75}$ days.
    This orbital period is reminiscent of the quasi-period of the strong dimming events observed by \cite{Rodriguez17}, and this third component may therefore be linked to this behaviour, as well as to a long-term RV modulation, which would explain the low velocities measured in 2017.

    To complete this study, we repeated the same analysis on both datasets assuming one companion only.
    This resulted in a  component of $\sim$188~M$_{jup}$, with an orbital period of about 247 days.
    Although the fit itself is as good as the two-companion hypothesis, the period found is not consistent with the apparent modulation of about 100 d found in the 2020-2021 dataset using a periodogram analysis.
    Furthermore, the minimum RV modulation expected for the epoch of the 2017 observations based on the derived orbital parameters  is still significantly higher than the values measured.
    The goodness of the fit is only reached thanks to a global decrease in the modulation induced by the stellar activity, which is unexpected.
    As both the $\sim$100-day modulation and the minimum at the epoch of the 2017 observations agree with the two-companion hypothesis, we favour this latter, although we cannot completely exclude the possibility that there is only one companion.

    We therefore conclude that there is at least one companion to HQ Tau.
    We also strongly suspect that this companion is a brown dwarf accompanied by a third component, which would explain the low velocities reported in Paper I.

\begin{acknowledgements}
    We thanks the anonymous referee for the very detailed and pertinent comments which drastically increased the robustness of this work.

    Based on observations obtained at the Canada–France– Hawaii Telescope (CFHT) which is operated from the summit of Maunakea by the National Research Council of Canada, the institut National des Sciences de l’Univers of the Centre National de la Recherche Scientifique of France, and the University of Hawaii. The observations at the Canada–France–Hawaii Telescope were performed with care and respect from the summit of Maunakea which is a significant cultural and historic site.

    Based on observations made at Observatoire de Haute Provence (CNRS), France.

    We thank the TBL team for providing service observing with Neo-Narval.

    We thank A. Carmona for his help in reducing the SPIRou observations, especially for the tellurics correction.
    
    This project has received funding from the European Research Council (ERC) under the European Union’s Horizon 2020 research and innovation programme (grant agreement No 742095 ; SPIDI : Star-Planets-Inner Disk- Interactions).
      
    This work has made use of the VALD database, operated at Uppsala University, the Institute of Astronomy RAS in Moscow, and the University of Vienna. 
\end{acknowledgements}

%-------------------------------------------------------------------

\bibliographystyle{aa}
\bibliography{bib}

%-------------------------------------------------------------------

\begin{appendix}

    % \section{Log of observations}
    \section{Additional tables}
    
    \label{ap:logobs}
    % In this section we present the journals of observation for the various instruments we used.
    % The SPIRou, ESPaDOnS, SOPHIE, and Neo-Narval observations are summarized in Table~\ref{tab:logspirou}, \ref{tab:logespadons}, \ref{tab:logsophie}, and \ref{tab:lognarval}, respectively.
    % The corresponding radial velocity and BIS measurement are summarized in Table~\ref{tab:vradCFHT}.% (see Sect.~\ref{subsec:RV} and \ref{subsec:LSD_BIS}).
    In this section we present the observation logs for the SOPHIE, Neo-Narval, SPIRou, and ESPaDOnS data sets in Table~\ref{tab:logsophie}, \ref{tab:lognarval}, \ref{tab:logspirou}, and \ref{tab:logespadons}, respectively.
    The corresponding RV and BIS measurements are summarised in Table~\ref{tab:vradCFHT}.% (see Sect.~\ref{subsec:RV} and \ref{subsec:LSD_BIS}).
    
        \begin{table}[H]
            
            \caption{Log of SOPHIE observations. }
            \centering
            \begin{tabular}{l l l l}
                \hline\hline
                Date & HJD & S/N & S/N$_I$ \\
                 & ($-$2\,450\,000 d) & & \\
                 \hline
                2020 Sep 03 & 9095.61 & 54 & 3748 \\
                2020 Sep 11 & 9103.58 & 50 & 3559 \\
                2020 Sep 17 & 9109.63 & 47 & 3592 \\
                2020 Oct 01 & 9123.6 & 61 & 3843 \\
                2020 Oct 10 & 9132.61 & 52 & 3816 \\
                2020 Oct 15 & 9137.62 & 47 & 3716 \\
                2020 Oct 24 & 9146.61 & 47 & 3755 \\
                2020 Oct 31 & 9153.53 & 60 & 3974 \\
                2020 Nov 07 & 9160.6 & 51 & 3733 \\
                2020 Nov 21 & 9174.66 & 40 & 3301 \\
                $^{\rm \dag}$2020 Nov 27 & 9181.5 & 51 & - \\
                2020 Dec 19 & 9202.52 & 44 & 3995 \\
                2021 Feb 03 & 9249.34 & 56 & 3906 \\
                2021 Feb 17 & 9263.36 & 64 & 3986 \\
        
                \hline
            \end{tabular}
            %\tablefoot{Namely, the date of observation, the HJD at the middle of exposure, the S/N at the order 31 (600 nm), and the S/N of the LSD Stokes I profile. The "-" sign means that the LSD Stokes I profile was not used in the analysis. The \dag\ symbol means that the observation is suspected to be contaminated by the moon and was thus not used in this work.}
            \tablefoot{Namely, the date of observation, the HJD at the middle of exposure, the S/N at the order 31 (600 nm), and the S/N of the LSD Stokes I profile. The "-" sign means that the LSD Stokes I profile was not used in the analysis. The \dag\ symbol indicates a moon contamination.}
    
            \label{tab:logsophie}
        \end{table}
    
        \begin{table}[H]
            \caption{Log of Neo-Narval observations. }
            \centering
            \begin{tabular}{l l l l}
                \hline\hline
                Date & HJD & S/N & S/N$_I$ \\
                 & ($-$2\,450\,00 d) & & \\
                 \hline
                2020 Sep 05 & 9097.63 & 20 & 1074 \\
                2020 Sep 13 & 9105.63 & 18 & 1698 \\
                2020 Sep 30 & 9122.65 & 7 & - \\
                2020 Oct 18 & 9140.54 & 7 & - \\
                2020 Oct 30 & 9152.58 & 10 & 1378 \\
                2020 Nov 22 & 9175.53 & 17 & - \\
                2020 Dec 15 & 9199.44 & 10 & - \\
                2020 Dec 22 & 9206.48 & 15 & 1952 \\
                2021 Jan 06 & 9221.37 & 14 & 1317 \\
                2021 Feb 10 & 9256.38 & 19 & 1640 \\
                2021 Feb 23 & 9269.34 & 8 & 806 \\
        
                \hline
            \end{tabular}
            %\tablefoot{Namely, the date of observation, the HJD at the middle of exposure, the S/N at 600 nm, and the S/N of the LSD Stokes I profile. The "-" sign means that the LSD Stokes I profile was not used in the analysis.}
            \tablefoot{Same as Table~\ref{tab:logsophie}.}
            \label{tab:lognarval}
        \end{table}
        
        \begin{table}[H]
            \caption{Log of SPIRou observations. }
            \centering
            \begin{tabular}{l l l l}
                \hline\hline
                Date & HJD & S/N & S/N$_I$ \\
                 & ($-$2\,450\,000 d) & & \\
                 \hline
                2020 Sep 04 & 9097.14 & 80 & 944 \\
                2020 Sep 06 & 9099.11 & 85 & 919 \\
                2020 Sep 21 & 9114.07 & 91 & 1020 \\
                2020 Sep 22 & 9115.1 & 84 & 790 \\
                2020 Sep 25 & 9118.07 & 82 & 1050 \\
                2020 Sep 27 & 9120.04 & 90 & 970 \\
                2020 Oct 04 & 9127.03 & 90 & 937 \\
                2020 Oct 05 & 9127.99 & 73 & 1033 \\
                2020 Oct 07 & 9129.99 & 88 & 1004 \\
                2021 Jan 05 & 9219.86 & 87 & 1173 \\
                    
                \hline
            \end{tabular}
            \tablefoot{Same as Table~\ref{tab:logsophie} with S/N given at the order 47.}
            \label{tab:logspirou}
        \end{table}
        
        \begin{table}[H]
            
            \caption{Log of ESPaDOnS observations. }
            \centering
            \begin{tabular}{l l l l}
                \hline\hline
                Date & HJD & S/N & S/N$_I$ \\
                (2020) & ($-$2\,450\,000 d) & & \\  
                \hline
                $^{\rm \dag}$Nov 30 & 9183.99 & 90 & - \\
                Dec 3 & 9187.09 & 73 & 2354 \\
                Dec 6 & 9190.0 & 82 & 2514 \\
                Dec 8 & 9191.92 & 81 & 2526 \\
                \hline
            \end{tabular}
            % \tablefoot{Namely, the date of observation, the HJD at the middle of exposure, the S/N at the order 31 (730 nm), and the S/N of the LSD Stokes I profile. The "-" sign means that the LSD Stokes I profile was not used in the analysis. The \dag\ symbol means that the observation is suspected to be contaminated by the moon and was thus not used in this work.}
            \tablefoot{Same as Table~\ref{tab:logspirou} with S/N given at the order 31 (730 nm).}
            \label{tab:logespadons}
        \end{table}

        \begin{table}[H]
                \centering
                \caption{Radial velocity and BIS measured in this work.}
                \begin{tabular}{l l l l l l}
                    \hline\hline
                    HJD & \vrad & $\delta$\vrad & BIS & $\delta$BIS & Instrument \\
                    \cline{2-5}
                    % (d) & (\kms) & (\kms) & (\kms) & (\kms) & \\
                    (d) & \multicolumn{4}{c}{(\kms)} & \\
    
                    \hline
                    9097.14 & 18.92 & 0.30 & $-$4.48 & 1.10 & SPIRou \\
                    9099.11 & 18.05 & 0.32 & $-$4.23 & 1.76 & SPIRou \\
                    9114.07 & 18.01 & 0.35 & $-$3.51 & 0.74 & SPIRou \\
                    9115.10 & 16.71 & 1.43 & $-$0.53 & 1.44 & SPIRou \\
                    9118.07 & 15.77 & 0.53 & 0.54 & 0.72 & SPIRou \\
                    9120.04 & 16.99 & 0.14 & 0.81 & 0.80 & SPIRou \\
                    9127.03 & 16.57 & 0.13 & $-$0.13 & 1.07 & SPIRou \\
                    9127.99 & 15.91 & 0.45 & 0.29 & 0.73 & SPIRou \\
                    9129.99 & 16.02 & 0.77 & $-$0.21 & 0.98 & SPIRou \\
                    9219.86 & 12.74 & 0.42 & 1.82 & 3.66 & SPIRou \\
                    &&&&& \\
                    9187.09 & 20.08 & 0.18 & 0.42 & 1.52 & ESPaDOnS \\
                    9190.00 & 20.31 & 0.46 & $-$1.49 & 1.32 & ESPaDOnS \\
                    9191.92 & 17.52 & 0.55 & 0.28 & 0.98 & ESPaDOnS \\
                    &&&&& \\
                    9095.61 & 16.06 & 0.35 & $-$0.13 & 0.58 & SOPHIE \\
                    9103.58 & 16.12 & 0.29 & 0.89 & 1.29 & SOPHIE \\
                    9109.63 & 19.82 & 0.07 & $-$2.10 & 1.90 & SOPHIE \\
                    9123.60 & 19.46 & 0.29 & $-$2.21 & 2.01 & SOPHIE \\
                    9132.61 & 14.20 & 0.12 & 2.54 & 1.32 & SOPHIE \\
                    9137.62 & 11.77 & 0.56 & 3.70 & 1.85 & SOPHIE \\
                    9146.61 & 15.52 & 0.77 & $-$0.06 & 0.45 & SOPHIE \\
                    9153.53 & 20.98 & 0.17 & $-$2.58 & 2.34 & SOPHIE \\
                    9160.60 & 19.71 & 0.34 & $-$2.11 & 1.41 & SOPHIE \\
                    9174.66 & 20.95 & 0.52 & $-$2.80 & 2.65 & SOPHIE \\
                    9202.52 & 18.48 & 0.45 & $-$0.62 & 0.22 & SOPHIE \\
                    9249.34 & 12.81 & 0.67 & $-$0.79 & 4.29 & SOPHIE \\
                    9263.36 & 17.08 & 0.41 & 0.07 & 0.36 & SOPHIE \\
                    &&&&& \\
                    9097.63 & 21.41 & 0.83 & $-$4.48 & 1.10 & Neo-Narval \\
                    9105.63 & 13.74 & 0.60 & $-$0.62 & 1.89 & Neo-Narval \\
                    9122.65 & 15.93 & 0.08 & - & - & Neo-Narval \\
                    9140.54 & 14.09 & 1.31 & - & - & Neo-Narval \\
                    9152.58 & 21.45 & 1.14 & $-$2.83 & 2.94 & Neo-Narval \\
                    9175.53 & 17.61 & 0.50 & - & - & Neo-Narval \\
                    9199.44 & 24.15 & 0.21 & - & - & Neo-Narval \\
                    9206.48 & 18.78 & 0.27 & 0.27 & 1.91 & Neo-Narval \\
                    9221.37 & 16.38 & 0.83 & $-$0.911 & 2.66 & Neo-Narval \\
                    9256.38 & 10.14 & 0.45 & 2.90 & 2.51 & Neo-Narval \\
                    9269.34 & 18.59 & 1.22 & - & - & Neo-Narval \\
                \hline\end{tabular}
                \tablefoot{The columns are listing the days of observation ($-$2\,450\,000~days), the radial velocity, its uncertainty, the BIS, its uncertainty, and the instrument of each observation used in this work.}
                \label{tab:vradCFHT}
            \end{table}
    \FloatBarrier
    %\newpage
    \section{Multi-variate GPR assuming the stellar activity only}
    \label{ap:GPonly}
    
    Here we present the results of the multi-variate GPR analysis considering that the variations are induced by the stellar activity only. We are thus fitting only the GP, without the Keplerian modulation.
    
        \begin{figure*}
            \centering
            \sidecaption
            \includegraphics[width=.70\textwidth]{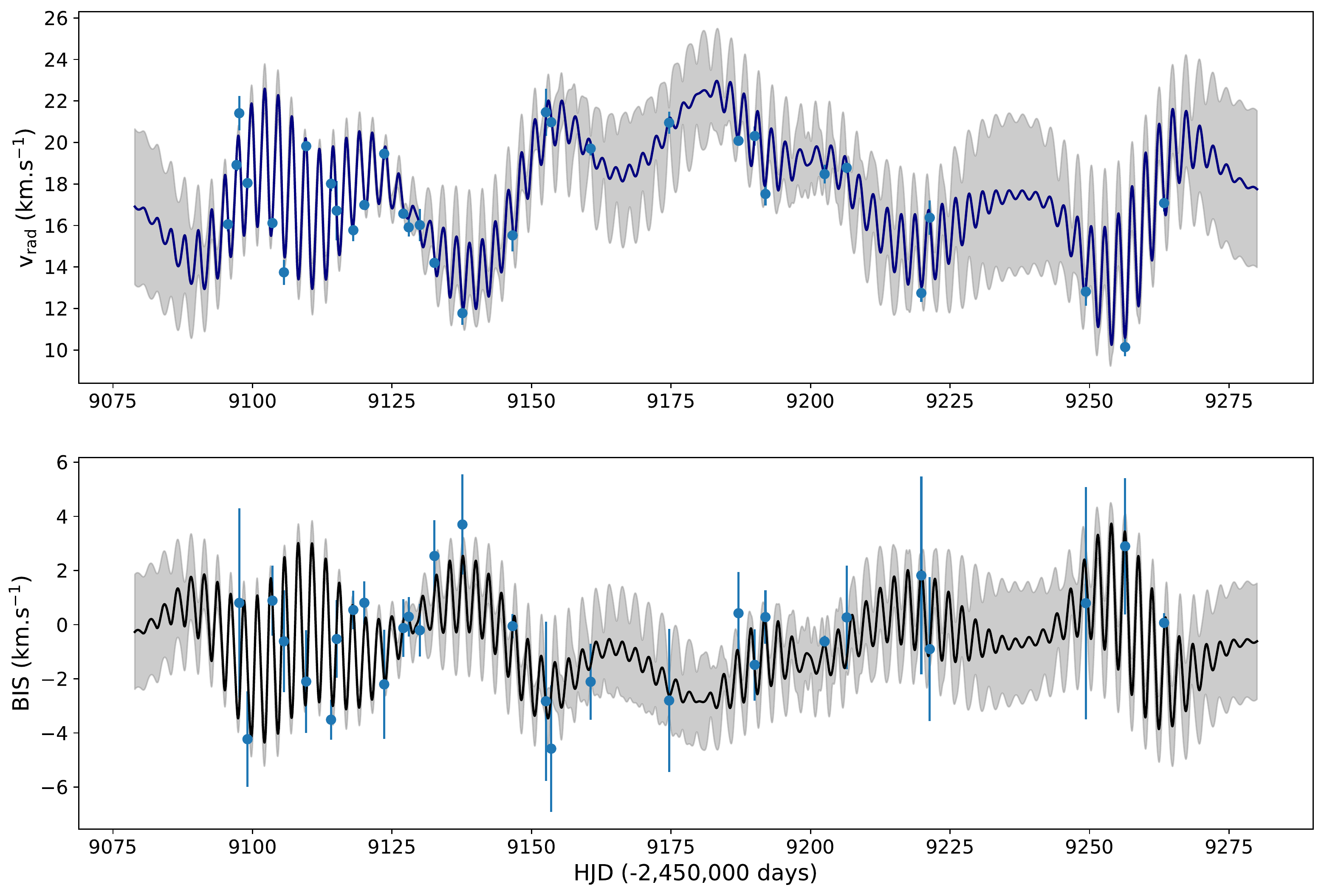}
            \caption{Fit of the RV \textit{(top)} and BIS \textit{(bottom)} curves with \texttt{pyaneti} assuming stellar activity only. The blue points are the measurements with their uncertainties, and the black curve is the GP fit of the stellar activity. The grey-shaded area is the 1$\sigma$ uncertainty of the fit.}
            \label{fig:RV200p}
        \end{figure*}
    
        \begin{table}[H]
            \centering
            \caption{Same as Table~\ref{tab:GPresults}, assuming the GP only.}
            \begin{tabular}{l c c}
            \hline\hline
                Parameter & Prior & Value \\
                \hline
                
                Fitted parameters & & \\
                A$_1$ (\kms) & $\mathcal{U}$[$-$20.0, 0.0] & $-$3.21$^{+1.14}_{-1.40}$ \\
                B$_1$ (\kms) & $\mathcal{U}$[$-$20.0, 0.0] & $-$6.45$^{+3.25}_{-5.33}$ \\
                A$_2$ (\kms) & $\mathcal{U}$[0.0, 20.0] & 1.27$^{+0.93}_{-0.75}$ \\
                B$_2$ (\kms) & $\mathcal{U}$[0.0, 20.0] & 5.55$^{+4.29}_{-2.78}$ \\
                $\lambda_e$ (d) & $\mathcal{U}$[1.,30] & 7.54$^{+1.68}_{-2.02}$ \\
                $\lambda_p$ & $\mathcal{U}$[0.1,10] & 4.57$^{+3.18}_{-2.06}$ \\
                P$_{GP}$ (d) & $\mathcal{N}$[2.42, 0.5] & 2.43$^{+0.08}_{-0.06}$ \\
                \hline
                %&&\\
                Derived parameters & & \\
                RV$_{syst}$ (\kms) &  & 17.3$^{+1.1}_{-1.1}$ \\
                %&&\\
                \hline
            \end{tabular}
            
            \label{tab:GPresultsGPonly}
        \end{table}

    \FloatBarrier
    
    \section{Multi-variate GPR on 2017 and 2020 radial velocity measurements}
    \label{ap:GPR1720}
    
\begin{figure*}
            \centering
            \sidecaption
            \includegraphics[width=.70\textwidth]{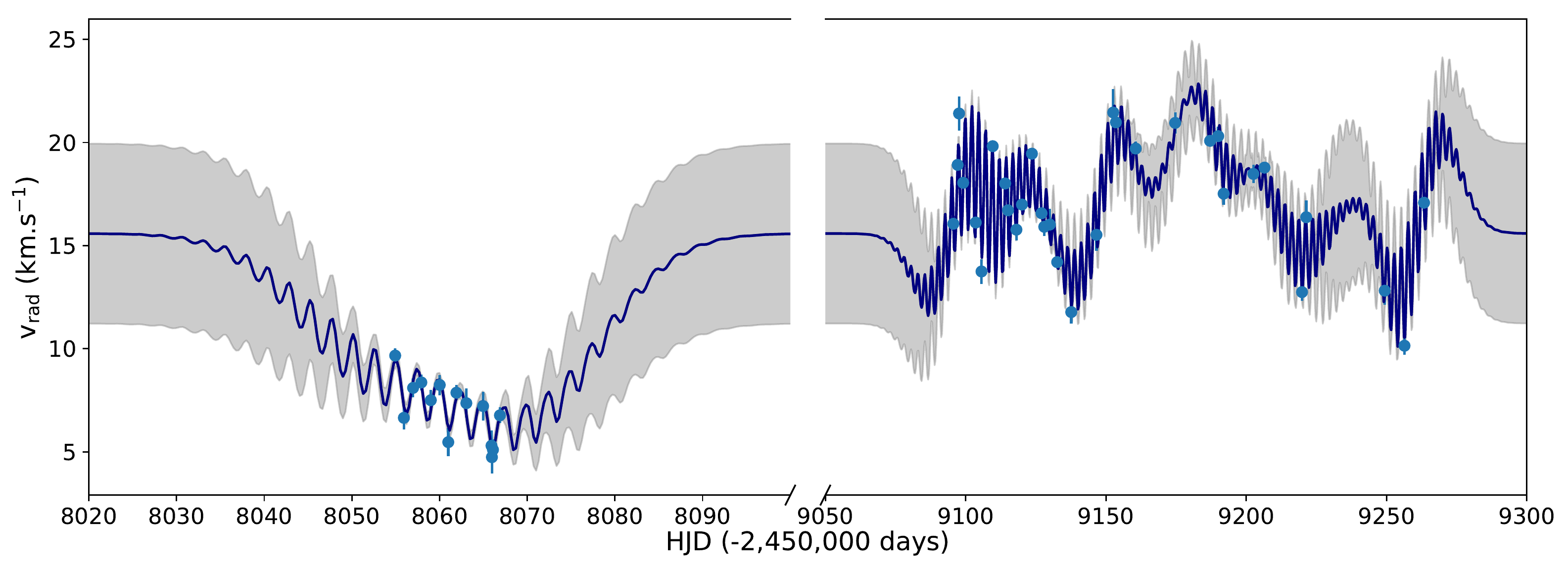}
            \caption{Joint fit of the 2017 and 2020 RV curve with \texttt{pyaneti} assuming stellar activity only. The blue points are the measurements with their uncertainties, the black curve is the GP fit of the stellar activity. The grey-shaded area is the 1$\sigma$ uncertainty of the fit.}
            \label{fig:RV17200p}
        \end{figure*}

\begin{figure*}
            \centering
            \sidecaption
            \includegraphics[width=.70\textwidth]{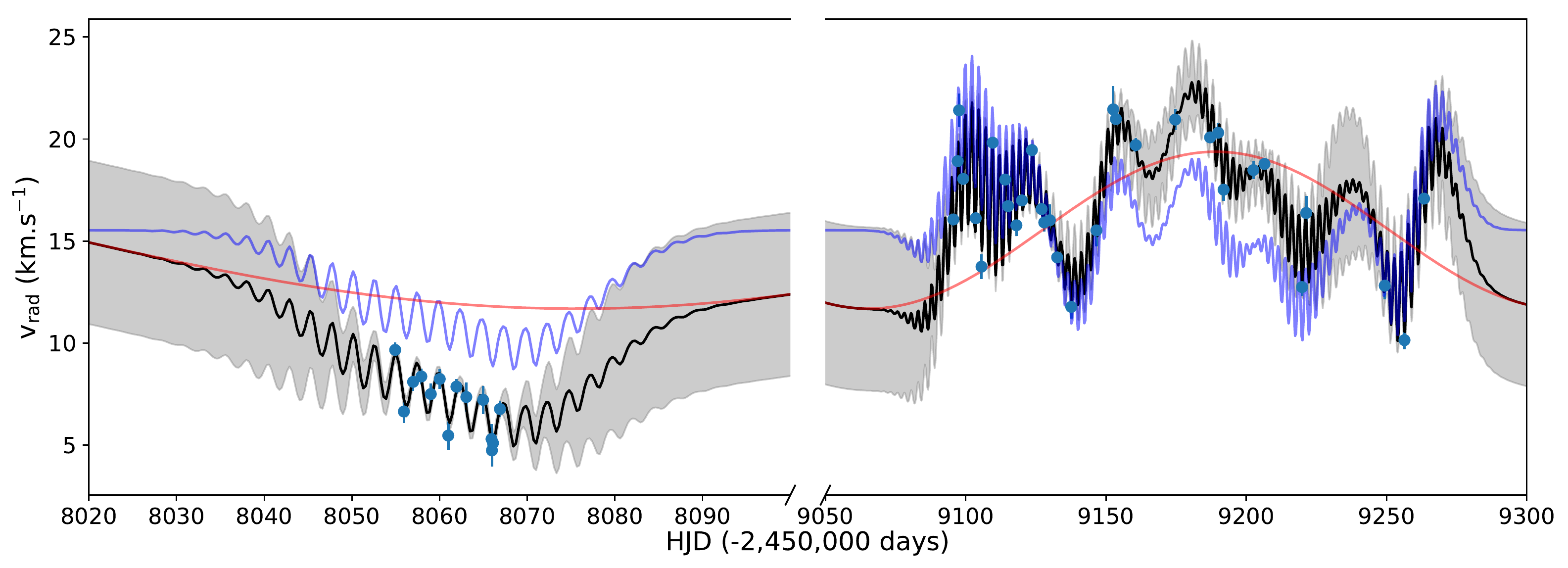}
            \caption{Joint fit of the 2017 and 2020 RV curve with \texttt{pyaneti} assuming one companion. The blue points are the measurements with their uncertainties, the black curve is the GP fit of the stellar activity (purple) plus the Keplerian modulation by the companion (red). The grey-shaded area is the 1$\sigma$ uncertainty of the fit. The x-axis has been cut for more visibility.}
            \label{fig:RV17201p}
        \end{figure*}

\begin{figure*}
            \centering
            \sidecaption
            \includegraphics[width=.70\textwidth]{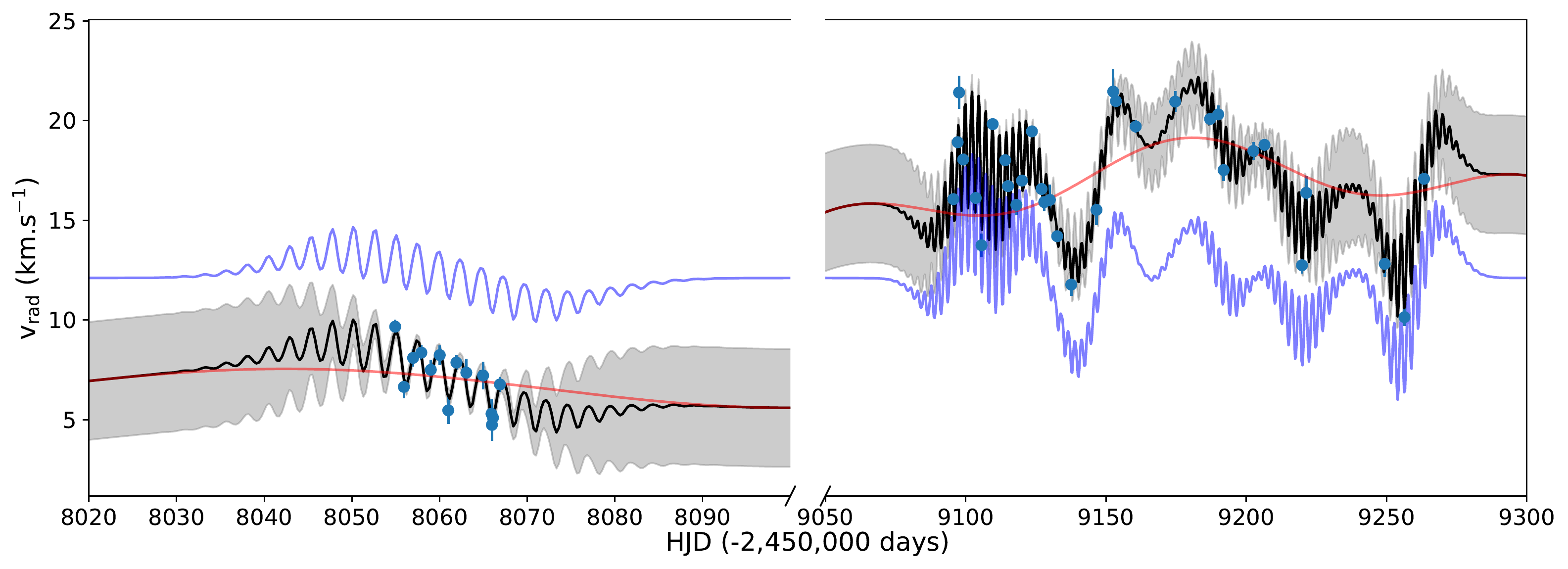}
            \caption{Joint fit of the 2017 and 2020 RV curve with \texttt{pyaneti} assuming two companions. The blue points are the measurements with their uncertainties, the black curve is the GP fit of the stellar activity (purple) plus the two Keplerian modulation by the two companions (red). The grey-shaded area is the 1$\sigma$ uncertainty of the fit. The x-axis has been cut for more visibility.}
            \label{fig:RV17202p}
        \end{figure*}

    We present here the results of the analysis of the combined 2017 and 2020 datasets of HQ Tau discussed in Sect.~\ref{sec:discussion}.
    We first present the results obtained assuming the stellar activity only (Table~\ref{tab:GPresults1720GPonly} and Fig.~\ref{fig:RV17200p}).
    The results of the one-companion hypothesis are shown in Fig.~\ref{fig:RV17201p} and summarized in Table~\ref{tab:GPresults17201p}.
    The similar figure and table for the two-companion hypothesis are presented in Fig.~\ref{fig:RV17202p} and Table~\ref{tab:GPresults17202p}.
    The two hypothesis are discussed in Sect.~\ref{sec:discussion}.

        \begin{table}[H]
            \centering
            \caption{Same as Table~\ref{tab:GPresults}, but for the fit of the joint 2017 and 2020 data sets, assuming the GP only.}
            \begin{tabular}{l c c}
            \hline\hline
                Parameter & Prior & Value \\
                \hline
                
                Fitted parameters & & \\
                A$_1$ (\kms) & $\mathcal{U}$[$-$20.0, 0.0] & $-$4.14$^{+7.79}_{-1.19}$ \\
                B$_1$ (\kms) & $\mathcal{U}$[$-$20.0, 0.0] & $-$3.39$^{+8.18}_{-1.78}$ \\
                A$_2$ (\kms) & $\mathcal{U}$[0.0, 20.0] & 0.23$^{+0.23}_{-0.39}$ \\
                B$_2$ (\kms) & $\mathcal{U}$[0.0, 20.0] & 2.81$^{+1.43}_{-6.72}$ \\
                $\lambda_e$ (d) & $\mathcal{U}$[1.,30] & 9.09$^{+1.84}_{-1.33}$ \\
                $\lambda_p$ & $\mathcal{U}$[0.1,10] & 3.26$^{+0.99}_{-0.82}$ \\
                P$_{GP}$ (d) & $\mathcal{N}$[2.42, 0.5] & 2.44$^{+0.06}_{-0.06}$ \\
                \hline
                %&&\\
                Derived parameters & & \\
                RV$_{syst}$ (\kms) &  & 15.6$^{+1.4}_{-1.5}$ \\
                %&&\\
                \hline
            \end{tabular}
            
            \label{tab:GPresults1720GPonly}
        \end{table}
    
        \begin{table}[H]
            \centering
            \caption{Same as Table~\ref{tab:GPresults}, but for the fit of the joint 2017 and 2020 data sets assuming one companion.}
            \begin{tabular}{l c c}
            \hline\hline
                Parameter & Prior & Value \\
                \hline
                
                Fitted parameters & & \\
                T0 (HJD$-$2\,450\,000)& $\mathcal{U}$[8900,9200]  & 9003.55$^{+51.64}_{-20.67}$ \\
                P$_{orb}$ (d) & $\mathcal{U}$[0,300] & 247.4$^{+10.7}_{-50.2}$ \\
                e & $\mathcal{F}$[0] &  \\
                $\omega$ & $\mathcal{F}$[0] &  \\
                K (\kms) & $\mathcal{U}$[$-$10.0, 10.0] & 3.85$^{+1.65}_{-2.02}$ \\
                A$_1$ (\kms) & $\mathcal{U}$[$-$10.0, 5.0] & $-$3.83$^{+1.59}_{-1.22}$ \\
                B$_1$ (\kms) & $\mathcal{U}$[$-$10.0,10.0] & $-$2.66$^{+7.37}_{-2.02}$ \\
                A$_2$ (\kms) & $\mathcal{U}$[$-$10.0,10.0] & 0.39$^{+0.32}_{-0.44}$ \\
                B$_2$ (\kms) & $\mathcal{U}$[$-$5.0,5.0] & 2.31$^{+1.63}_{-6.21}$ \\
                $\lambda_e$ (d) & $\mathcal{U}$[1.,30] & 8.77$^{+1.76}_{-1.34}$ \\
                $\lambda_p$ & $\mathcal{U}$[0.1,5] & 3.08$^{+1.08}_{-0.86}$ \\
                P$_{GP}$ (d) & $\mathcal{N}$[2.42, 0.05] & 2.43$^{+0.04}_{-0.04}$ \\
                \hline
                %&&\\
                Derived parameters & & \\
                RV$_{syst}$ (\kms) &  & 15.5$^{+1.5}_{-1.5}$ \\
                %BIS$_{syst}$ (\kms) &  & -0.6$^{+0.3}_{-0.3}$ \\
                M$_b$ (M$_{jup}$) &  & 187.5$^{+95.4}_{-103.6}$ \\
                T$_{peri}$ (HJD$-$2\,450\,000) &  & 9064.38$^{+44.13}_{-20.18}$ \\
                a (AU) & & 0.98$^{+0.08}_{-0.24}$ \\
                %&&\\
                \hline
            \end{tabular}
            
            \label{tab:GPresults17201p}
        \end{table}
     
        \begin{table}[H]
            \centering
            \caption{Same as Table~\ref{tab:GPresults17201p} assuming two companions.}
            \begin{tabular}{l c c}
            \hline
                Parameter & Prior & Value \\
                \hline\hline
                
                Fitted parameters & & \\
                T0$_B$ (HJD$-$2\,450\,000) & $\mathcal{N}$[9095,40]  & 9084.22$^{+25.43}_{-25.50}$ \\
                P$_{orb,B}$ (d) & $\mathcal{N}$[100,50] & 126.2$^{+27.3}_{-22.5}$ \\
                e$_B$ & $\mathcal{F}$[0] &  \\
                $\omega_B$ & $\mathcal{F}$[0] &  \\
                K$_B$ (\kms) & $\mathcal{U}$[$-$10.0, 10.0] & 1.27$^{+0.7}_{-0.7}$ \\
                T0$_C$ (HJD$-$2\,450\,000) & $\mathcal{N}$[9275,100]  & 9389.71$^{+42.58}_{-39.44}$ \\
                P$_{orb,C}$ (d) & $\mathcal{U}$[500,1000] & 766.8$^{+91.0}_{-75.3}$ \\
                e$_C$ & $\mathcal{F}$[0] &  \\
                $\omega_C$ & $\mathcal{F}$[0] &  \\
                K$_C$ (\kms) & $\mathcal{U}$[0.0, 10.0] & 5.83$^{+1.82}_{-1.10}$ \\
                A$_1$ (\kms) & $\mathcal{U}$[$-$10.0,0.0] & $-$2.72$^{+4.50}_{-1.18}$ \\
                B$_1$ (\kms) & $\mathcal{U}$[$-$10.0,10.0] & $-$2.42$^{+6.00}_{-2.96}$ \\
                A$_2$ (\kms) & $\mathcal{U}$[$-$10.0,10.0] & 0.93$^{+0.86}_{-1.96}$ \\
                B$_2$ (\kms) & $\mathcal{U}$[$-$5.0,5.0] & 2.12$^{+1.65}_{-4.94}$ \\
                $\lambda_e$ (d) & $\mathcal{U}$[1.,30] & 7.96$^{+1.50}_{-1.37}$ \\
                $\lambda_p$ & $\mathcal{U}$[0.1,5] & 2.78$^{+1.23}_{-1.01}$ \\
                P$_{GP}$ (d) & $\mathcal{N}$[2.42, 0.05] & 2.44$^{+0.04}_{-0.04}$ \\
                \hline
                %&&\\
                Derived parameters & & \\
                RV$_{syst}$ (\kms) &  & 12.1$^{+1.4}_{-1.8}$ \\
                %BIS$_{syst}$ (\kms) &  & -0.4$^{+0.5}_{-0.4}$ \\
                M$_B$ (M$_{jup}$) &  & 47.7$^{+28.7}_{-24.9}$ \\
                T$_{peri,B}$ (HJD$-$2\,450\,000) &  & 9115.49$^{+24.86}_{-22.28}$ \\
                a$_{B}$ (AU) & & 0.61$^{+0.06}_{-0.13}$ \\
                M$_C$ (M$_{jup}$) &  & 464.8$^{+186.2}_{-105.6}$ \\
                T$_{peri,C}$ (HJD$-$2\,450\,000) &  & 9581.24$^{+61.13}_{-54.59}$ \\
                a$_{C}$ (AU) & & 2.16$^{+0.29}_{-0.34}$ \\
                %&&\\
                \hline
            \end{tabular}
            
            \label{tab:GPresults17202p}
        \end{table}
        
    \FloatBarrier
    \section{Choice of the priors}
    
    In this section we present how we choose the priors indicated in Tables \ref{tab:GPresults}, \ref{tab:GPresultsGPonly}, \ref{tab:GPresults1720GPonly}, \ref{tab:GPresults17201p}, and \ref{tab:GPresults17202p}.
    We are focusing this section on the Keplerian parameters for which we choose a normal distribution.
    
    \subsection{2020 dataset}
    \subsubsection*{Orbital period P$_{\rm orb}$}
    
    The Lomb-Scargle periodogram (Fig.~\ref{fig:RVHQTau}) pointed out an orbital period at 111 days, given the low peak and high FAP, we used only 1 significant digit, meaning 100 days. 
    Furthermore, the apparent modulation of the RV curve is consistent with such a period.
    Finally, from our first work on this object, we suspected an orbital period of the same order. Given these, the probability of having an orbital period of around 100 days seems higher than 50 or 150 days, explaining why we used a normal distribution instead of a linear one. 
    We choose a $\sigma$ of 50 to reach 0 days at 2-sigma and allow to exceed the 170 days of observation available.
    
    \subsubsection*{Transit epoch T0}
    
    Assuming that the low modulation of the RV curve in Fig.~\ref{fig:RVHQTau} is an orbital modulation, we estimated the time at periastron around HJD 2\,459\,170, from which we removed the estimated 100-day orbital period to avoid negative cycles. 
    Passing from the time at periastron to the transit epoch in our simple configuration means adding one-fourth of the period, meaning T0 = HJD 2\,459\,095. 
    Here again, this value is motivated by previous measurements/estimates, explaining why we used a normal distribution. 
    As T0 is modulo the orbital period, we choose the sigma to be lower than half the orbital period to avoid a misinterpretation of the posterior between T0 and T0+P$_{\rm orb}$.
    
    \subsection{Joint 2017 and 2020 datasets}
    \subsubsection*{2-companion model}
    
    For this hypothesis, the B component is supposed to be the same as the one modulation the 2020 dataset only, we thus kept the same priors for this component.
    Concerning the orbital motion of the companion C, we did not have any indication, justifying the use of a uniform distribution for the orbital period. 
    We used the approximate range of occurrence of the dimming events reported by \citet{Rodriguez17} of about 750 days, which is not reported to be strictly periodic, that we broadened symmetrically to reach the 1000 days reported between two of these events. 
    For the T0 priors, we first ran an MCMC with a uniform distribution, then we set the prior with the normal distribution around one of the peaks obtained to avoid multiple peaks in the posterior.

    \end{appendix}

%-------------------------------------------------------------------

\end{document}